\documentclass[aps, prl, superscriptaddress, citeautoscript, showpacs, twocolumn, amsmath,amssymb,aps]{revtex4-1}
\usepackage{float}
\usepackage[pdftex]{graphicx}
\usepackage{epstopdf}
\usepackage{amsmath}
\usepackage{subfigure}
\usepackage{fixltx2e}
\usepackage{dcolumn}
\usepackage{gensymb}
\usepackage{graphicx}
\usepackage{dcolumn}
\usepackage{bm}

\begin{document}
\title{CsV$_3$Sb$_5$: a $\mathbb{Z}_2$ topological kagome metal with a superconducting ground state}

\author{Brenden R. Ortiz}
 \email{ortiz.brendenr@gmail.com}
 \affiliation{Materials Department and California Nanosystems Institute, University of California Santa Barbara, Santa Barbara, CA, 93106, United States}%
 
\author{Samuel M.L. Teicher}
 \affiliation{Materials Department and California Nanosystems Institute, University of California Santa Barbara, Santa Barbara, CA, 93106, United States}%
 
\author{Yong Hu}%
 \affiliation{Hefei National Laboratory for Physical Sciences at the Microscale, Department of Physics and CAS Key Laboratory of Strongly-coupled Quantum Matter Physics, University of Science and Technology of China, Hefei, Anhui 230026, China}%

 \author{Julia L. Zuo}
 \affiliation{Materials Department and California Nanosystems Institute, University of California Santa Barbara, Santa Barbara, CA, 93106, United States}%

\author{Paul M. Sarte}
 \affiliation{Materials Department and California Nanosystems Institute, University of California Santa Barbara, Santa Barbara, CA, 93106, United States}%
 
 \author{Emily C. Schueller}
 \affiliation{Materials Department and California Nanosystems Institute, University of California Santa Barbara, Santa Barbara, CA, 93106, United States}%
 
 \author{A.M. Milinda Abeykoon}
 \affiliation{National Synchrotron Light Source II, Brookhaven National Laboratory, Upton, New York 11973, USA}%

 \author{Matthew J. Krogstad}
 \affiliation{Materials Science Division, Argonne National Laboratory, Argonne, Illinois 60439-4845, USA}%
 
 \author{Stefan Rosenkranz}
 \affiliation{Materials Science Division, Argonne National Laboratory, Argonne, Illinois 60439-4845, USA}%
 
 \author{Raymond Osborn}
 \affiliation{Materials Science Division, Argonne National Laboratory, Argonne, Illinois 60439-4845, USA}%
 
 \author{Ram Seshadri}
 \affiliation{Materials Department and California Nanosystems Institute, University of California Santa Barbara, Santa Barbara, CA, 93106, United States}%

  \author{Leon Balents}
  \affiliation{Kavli Institute for Theoretical Physics, University of California, Santa Barbara, Santa Barbara, California 93106, USA}
  
 \author{Junfeng He}
 \affiliation{Hefei National Laboratory for Physical Sciences at the Microscale, Department of Physics and CAS Key Laboratory of Strongly-coupled Quantum Matter Physics, University of Science and Technology of China, Hefei, Anhui 230026, China}%

 \author{Stephen D. Wilson}
 \email{stephendwilson@ucsb.edu}
 \affiliation{Materials Department and California Nanosystems Institute, University of California Santa Barbara, Santa Barbara, CA, 93106, United States}%

\date{\today}

\begin{abstract}

Recently discovered alongside its sister compounds KV$_3$Sb$_5$ and RbV$_3$Sb$_5$, CsV$_3$Sb$_5$ crystallizes with an ideal kagome network of vanadium and antimonene layers separated by alkali metal ions. This work presents the electronic properties of CsV$_3$Sb$_5$, demonstrating bulk superconductivity in single crystals with a T$_{c} = 2.5$\,K. The normal state electronic structure is studied via angle-resolved photoemission spectroscopy (ARPES) and density functional theory (DFT), which categorize CsV$_3$Sb$_5$ as a $\mathbb{Z}_2$ topological metal. Multiple protected Dirac crossings are predicted in close proximity to the Fermi level ($E_F$), and signatures of normal state correlation effects are also suggested by a high temperature charge density wave-like instability. The implications for the formation of unconventional superconductivity in this material are discussed. 
\end{abstract}

\maketitle

Kagome metals are a rich frontier for the stabilization of novel correlated and topological electronic states. Depending on the degree of electron filling within the kagome lattice, a wide array of instabilities are possible, ranging from bond density wave order \cite{PhysRevB.87.115135,PhysRevLett.97.147202}, charge fractionalization \cite{PhysRevB.81.235115, PhysRevB.83.165118}, spin liquid states \cite{yan2011spin}, charge density waves \cite{PhysRevB.80.113102}, and superconductivity \cite{PhysRevB.87.115135,ko2009doped}. Additionally, the kagome structural motif imparts the possibility of topologically nontrivial electronic  structures, where the coexistence of Dirac cones and flatbands promoting strong correlation effects may engender correlated topological states. For instance, the presence of magnetic order \cite{kang2020dirac, ye2018massive, morali2019fermi} in kagome compounds has been noted to stabilize novel quantum anomalous Hall behaviors, and electron-electron interactions in certain scenarios are proposed to drive the formation of topological insulating phases \cite{wen2010interaction}. 

One widely sought electronic instability on a two-dimensional kagome lattice is the formation of a superconducting ground state.  Layered kagome metals that superconduct are rare, and the interplay between the nontrivial topology accessible via their electronic band structures and the formation of an intrinsic superconducting state makes this a particularly appealing space for realizing exotic ground
states and quasiparticles.  Unconventional superconductivity is predicted to emerge via nesting-driven interactions in heavily doped kagome lattices \cite{PhysRevB.85.144402}.  This mechanism, first pointed out in theories for doped graphene (which shares the hexagonal symmetry of the kagome
lattice)\cite{nandkishore2012interplay,nandkishore2012chiral}, relies upon scattering between saddle points of a band at the M points of the 2d Brillouin zone, which are relevant when the system possesses a nearly hexagonal Fermi surface proximate to a topological transition. Superconductivity potentially competes with a variety of other electronic instabilities at different fillings \cite{PhysRevLett.110.126405,wen2010interaction}.  Realizing superconductivity in a two-dimensional kagome material that avoids these competing instabilities remains an open challenge.

Recently a new family of layered kagome metals that crystallize in the \textit{A}V$_3$Sb$_5$ structure (\textit{A}: K, Rb, Cs) was reported \cite{ortiz2019new}.  These materials crystallize into the $P$6/$mmm$ space group, with a kagome network of vanadium cations coordinated by octahedra of Sb. The compounds are layered, with the kagome sheets separated by layers of the \textit{A}-site alkali metal ions (Fig. 1). Compounds across the series are high mobility, two-dimensional metals with signatures of correlation effects and potential electronically-driven symmetry breaking. Recent studies have further shown that one variant, KV$_3$Sb$_5$, is a Dirac semimetal with an extraordinarily large anomalous Hall effect in the absence of long-range magnetic order \cite{YangKV3Sb5science}. Remarkably little however remains known about this new class of kagome metals, particularly with regards to their capacity for hosting correlated topological states.   

Here we identify that CsV$_3$Sb$_5$, the heaviest member of the new kagome compounds, is a $\mathbb{Z}_2$ topological metal with a superconducting ground state. Angle-resolved photoemission spectroscopy (ARPES) measurements combined with density-functional theory (DFT) calculations reveal the presence of multiple Dirac points near the Fermi level and predict topologically protected surfaces states only 0.05\,eV above the Fermi level at the $M$-points. Furthermore, both ARPES and DFT observe hexagonal Fermi surfaces, consistent with close proximity of saddle points at $M$. Magnetization, heat capacity, and electrical resistivity measurements reveal the onset of superconductivity at $T_c = 2.5$\,K and further identify a higher temperature $T^*=94$ K transition suggestive of charge density wave order.  Our work establishes CsV$_3$Sb$_5$ as a novel exfoliable, kagome metal with a superconducting ground state and protected Dirac crossings close to $E_F$. 

Single crystals of CsV$_3$Sb$_5$ were synthesized via a self-flux growth method \cite{supplemental}. Magnetization measurements were performed using a Quantum Design SQUID Magnetometer (MPMS3) in vibrating-sample measurement mode (VSM), and resistivity and heat capacity measurements were performed using a Quantum Design Dynacool Physical Properties Measurement System (PPMS).  Angle-resolved photoemission (ARPES) measurements were obtained at the Stanford Synchrotron Radiation Lightsource (SSRL, beamline 5-2), a division of the SLAC National Accelerator Laboratory using 120\,eV photons with an energy resolution better than 20\,meV.  Temperature-dependent X-ray diffraction data was collected at Brookhaven National Laboratory (beamline 28-ID-1) and at the Advanced Photon Source at Argonne National Laboratory (Sector 6-ID-D). Rietveld refinements of temperature-dependent diffraction were performed using TOPAS Academic V6.\cite{coelho2018topas} Structure visualization was performed with the VESTA software package \cite{Momma2011}, and the electronic structure of CsV$_3$Sb$_5$ was calculated in VASP $v$5.4.4 \cite{Kresse1994,Kresse1996a,Kresse1996b} using projector-augmented wave (PAW) potentials \cite{Blochl1994a,Kresse1999} with details described in the supplemental materials \cite{supplemental}.

\begin{figure}
\includegraphics[width=\columnwidth]{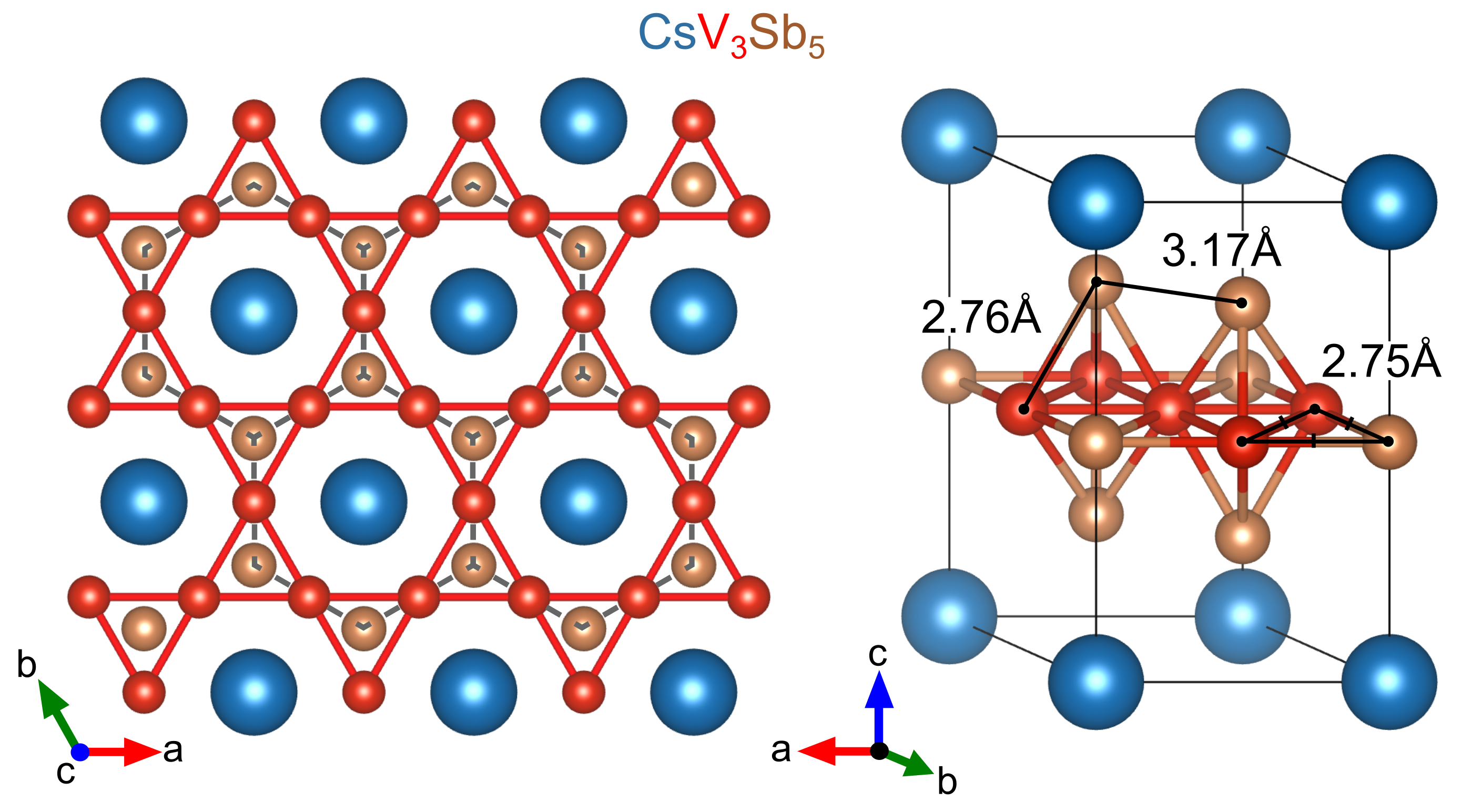}
\caption{CsV$_3$Sb$_5$ is a layered compound with a structurally perfect kagome network of vanadium. There are two distinct Sb sites in the structure: (1) a simple hexagonal net woven into the kagome layer, and (2) graphite-like layers of antimony (antimonene) above and below the kagome layer. All bonds $\leq 3.2$\AA\,have been drawn in the isometric perspective, highlighting the layered nature of CsV$_3$Sb$_5$}
\label{fig:Crystal}
\end{figure}

For an intuitive understanding of the CsV$_3$Sb$_5$ structure, we first consider the constituent sublattices. The hallmark two-dimensional kagome network is formed by the V1 sublattice, and is interpenetrated by a simple hexagonal net of Sb1 antimony. All interatomic distances within the kagome layer are equal (2.75\AA), as required by the high symmetry of the V1 (Wyckoff 3\textit{g}) and Sb1 (Wyckoff 1\textit{b}) sites. The Sb2 sublattice creates graphite-like layers of Sb (antimonene) that encapsulate the kagome sheets. The Cs1 sublattice naturally fills the space between the graphite-like sheets, and the nearest Cs-Sb distance is nearly 4\AA.

\begin{figure*}
  \includegraphics[width=\textwidth]{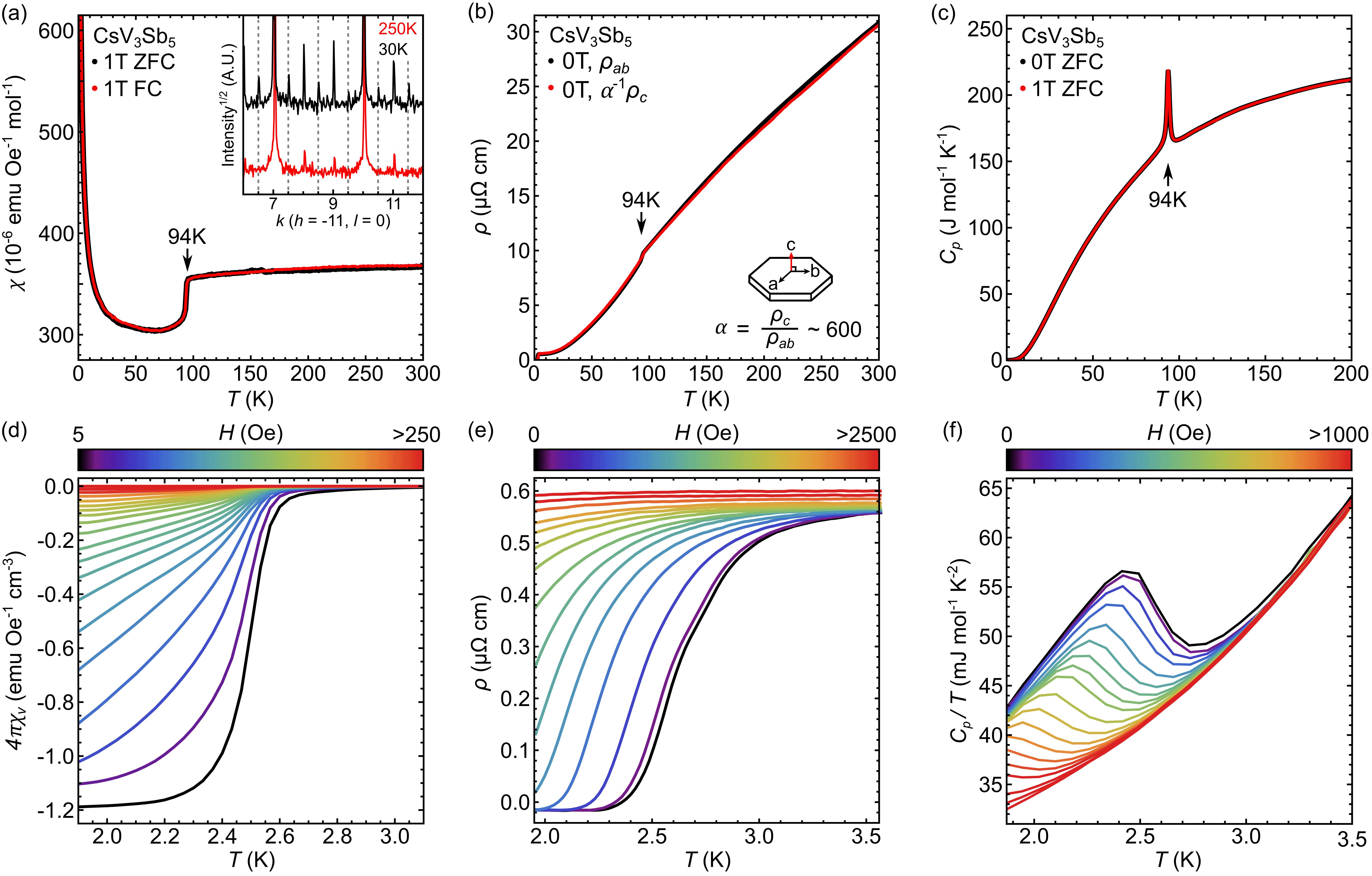}
  \caption{(a,c,e) Full temperature ranges for the magnetization, electrical resistivity, and heat capacity, respectively, shown for single crystals of CsV$_3$Sb$_5$. All measurements indicate presence of an anomaly $T^*$ at 94\,K, suspected to be an electronic instability (e.g. charge ordering). The inset in panel (a) shows linecuts through x-ray diffraction data below and above $T^*$.  Dashed lines denote the appearance of half-integer reflections (b, d, f) Field-dependent measurements at low temperatures, showing onset of superconductivity in magnetization, resistivity, and heat capacity, respectively. The $T_c$ for CsV$_3$Sb$_5$ is approximately 2.5\,K, with a slight suppression in resistivity due to high probe currents.}
\label{fig:Transport}
\end{figure*}

Bulk electronic properties of CsV$_3$Sb$_5$ were studied via electron transport, magnetization, and heat capacity measurements. Figures \ref{fig:Transport}(a, b, c) show characterization data collected across a broad range of temperatures.  Magnetization data collected under $\mu_0 H=1$ T are plotted as susceptibility $\chi=\frac{M}{H}$ in Fig. 2 (a) and show a high temperature response ($T>100$ K) consistent with Pauli paramagnetism. As a rough estimate, DFT calculations of the density of states at the Fermi level $g(E_F) \approx 10$\,eV$^{-1}$cell$^{-1}$ estimate $\chi \approx 200\times10^{-6}$\,emu\,Oe$^{-1}$mol$^{-1}$, which agrees reasonably well with the experimental data. 

At temperatures below 94\,K, a sharp drop in the magnetization data denotes the onset of a phase transition, noted as $T^*$. This transition also appears as an inflection point in the resistivity data shown in Figure \ref{fig:Transport}(b), where temperature-dependent resistivity data with current flowing both in the kagome planes ($\rho_ab$) and between the planes ($\rho_c$) are plotted.  The out-of-plane resistivity is nearly 600 times larger than in-plane, emphasizing the two-dimensional nature of the Fermi surface. Heat capacity data plotted in Fig. 2 (c) also illustrate a strong entropy anomaly at $T^*=94$ K.  The integrated entropy released through the $T^*$ transition is approximately $\Delta S = 1.6$\,J\,mol$^{-1}$K$^{-1}$, and is naively too small to account for collective spin freezing of free V-moments.  Instead, it likely arises from freezing within the charge sector \cite{wen2010interaction}, suggesting a potential charge/bond density wave anomaly that will be discussed later in this manuscript.

\begin{figure}
	\includegraphics[width=\columnwidth]{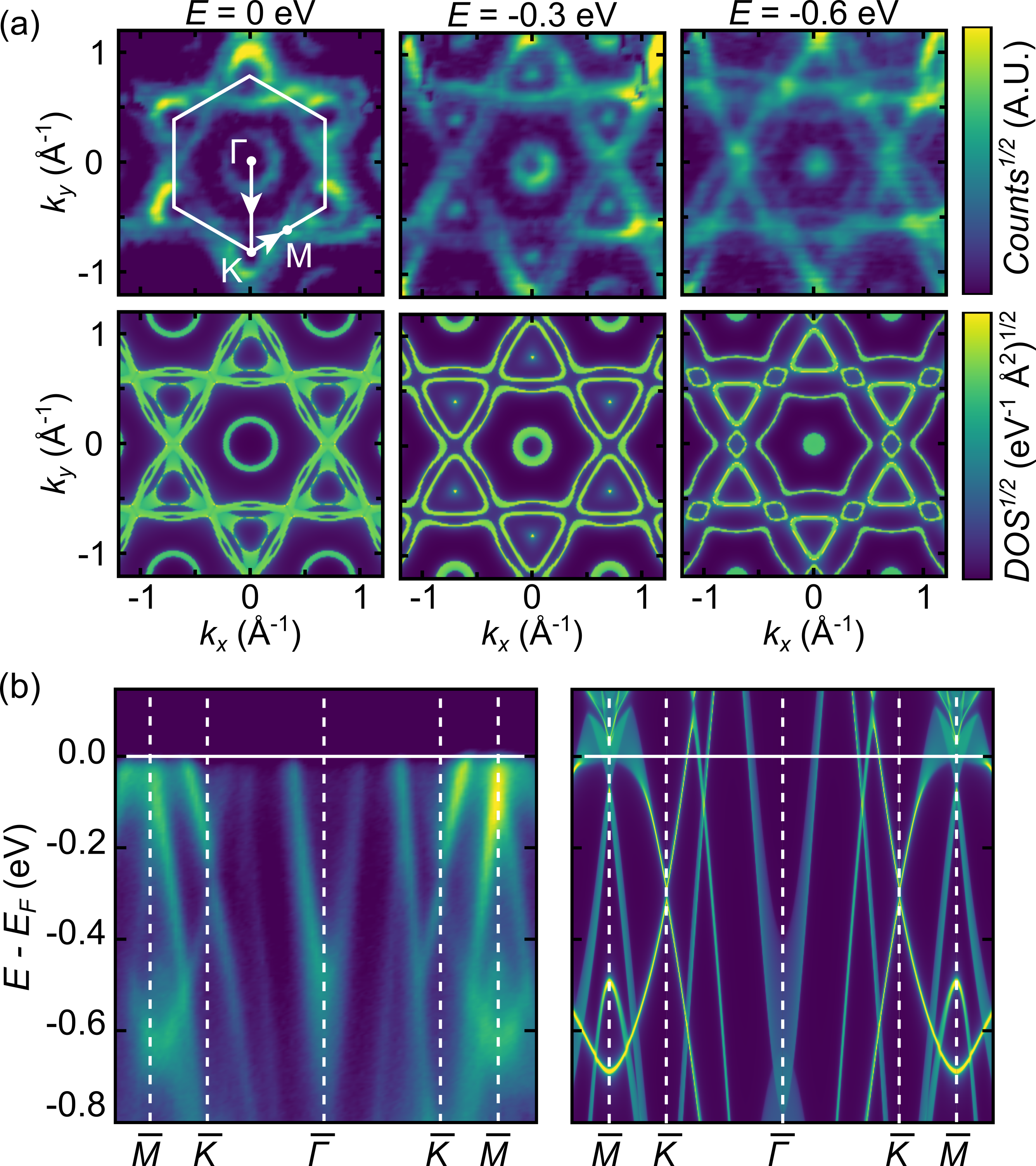}
	\caption{Experimental ARPES data and comparison with DFT calculations. (a) A selection of constant energy maps at 80\,K are compared with DFT calculations, showing excellent agreement. The hexagonal Brillouin zone is superimposed on the $E=0$\,eV data. (b) ARPES and DFT data tracing from $M$-$K$-$\Gamma$-$K$-$M$ reveal multiple Dirac points throughout the dispersion. Surface states can be observed in the DFT data at the $M$-point, slightly above $E_F$.}
	\label{fig:ARPES}
\end{figure}

Figures \ref{fig:Transport}(d-f) show the onset of superconductivity in magnetization, resistivity, and heat capacity, respectively. In all cases, the onset of superconductivity occurs at approximately $T_c = 2.5$\,K. Magnetization data reveal bulk superconductivity and a well-defined Meissner state, and heat capacity measurements show a sharp entropy anomaly at the superconducting transition, although, due to a limited temperature regime, we are unable to fully characterize the gapped behavior far below $T_c$.  The slight offset in the onset of $T_c$ in electrical resistivity (Figure \ref{fig:Transport}(e)) measurements is due to the high probe currents (8\,mA) used in the DC measurement. Reduced currents show $T_c$ return to nominal values, although the data quality suffers significantly due to the low resistivity of CsV$_3$Sb$_5$ single crystals.    

Having determined that CsV$_3$Sb$_5$ is a bulk kagome superconductor with a transition temperature $T_c = 2.5$\,K, we next examine the normal state metal via a combination of ARPES measurements and DFT modeling.  Figure \ref{fig:ARPES}(a) shows both ARPES and DFT modeling data with the hexagonal Brillouin zone superimposed on the $E=0$\,eV constant energy contour and high symmetry points $K$, $M$, and $\Gamma$ labeled. Data collected with differing photon energies did not reveal any appreciable dispersion along $k_z$, consistent with a quasi-2D band structure.  ARPES data were collected at 50\,K, 80\,K, 100\,K, and 120\,K, and no resolvable changes were observed in the band structure when transitioning through the $T^*$ transition. The DFT model shows remarkable agreement with the ARPES data, recovering all experimental observed crossings below the Fermi level. Figure \ref{fig:ARPES}(b) shows both the measured and calculated electronic structure hosts multiple Dirac points at finite binding energies.   

While inaccessible in the present ARPES data, the DFT model further reveals multiple topological band features slightly above the Fermi energy. The $\overline{M}$-point is of particular interest, as $M$ is a time-reversal invariant momentum (TRIM) point. Figure 4 shows the results of a tight binding calculation of surface states in CsV$_3$Sb$_5$, where bright spots slightly above the Fermi energy indicate surface states.

\begin{figure}
	\includegraphics[width=\columnwidth]{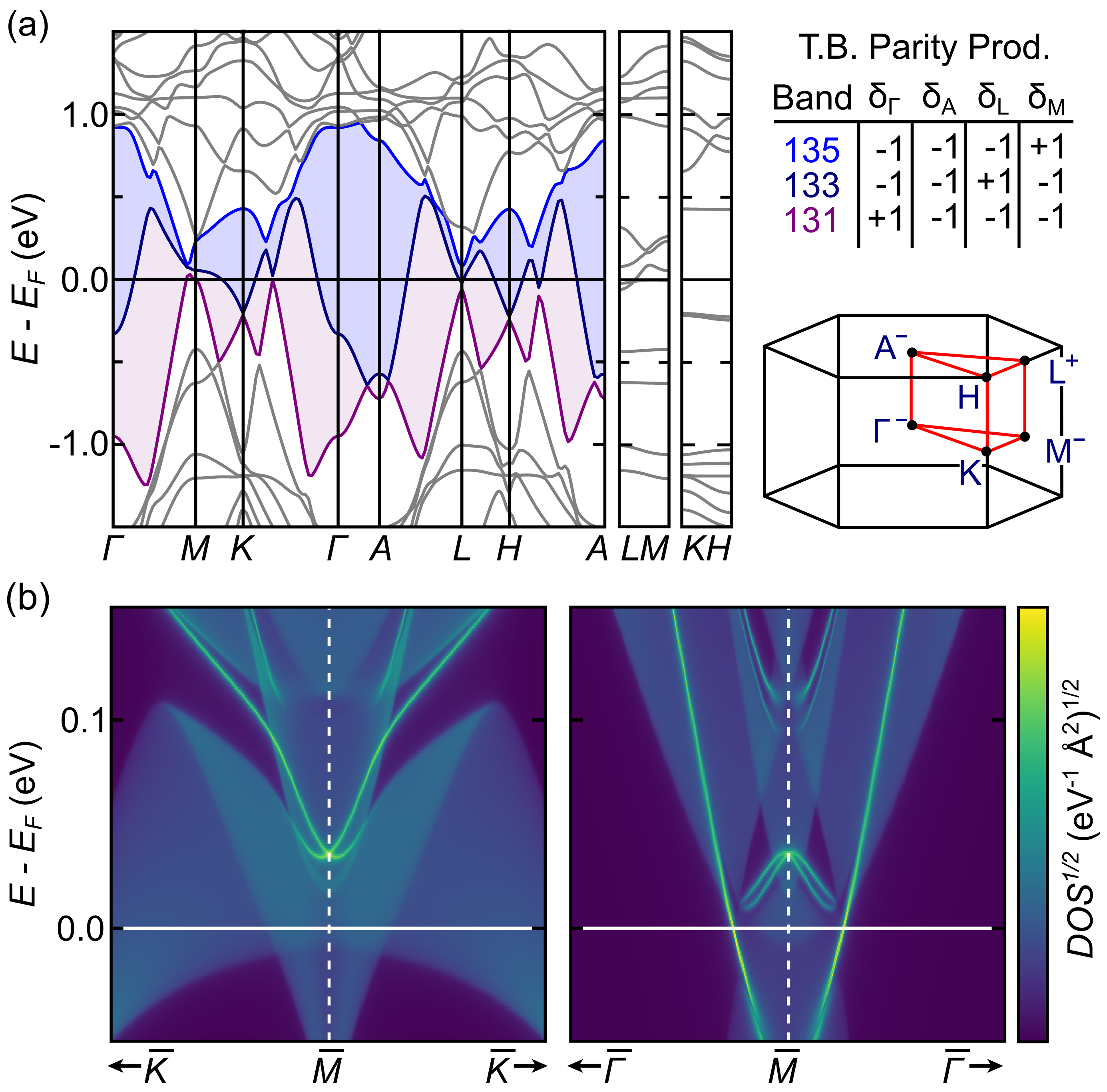}
	\caption{(a) Calculated band structure of CsV$_3$Sb$_5$ along high symmetry directions across the Brillouin zone.  A continuous direct gap (shaded) is noted and high symmetry points in the BZ are labeled. (b) Tight-binding model of CsV$_3$Sb$_5$ showing topologically-protected surfaces states that manifest at the time-reversal invariant momentum $\overline{M}$-point.}
	\label{fig:topo}
\end{figure}

Unlike many heavily studied kagome lattices (e.g. ZnCu$_3$(OH)$_6$Cl$_2$,\cite{braithwaite2004herbertsmithite,freedman2010site,han2012fractionalized} Fe$_3$Sn$_2$,\cite{ye2018massive,wang2020giant} Mn$_3$Ge,\cite{nayak2016large,kiyohara2016giant} Co$_3$Sn$_2$S$_2$\cite{vaqueiro2009powder,xu2018topological,morali2019fermi}), CsV$_3$Sb$_5$ does not exhibit resolvable magnetic order. Given that CsV$_3$Sb$_5$ possesses both time-reversal and inversion symmetry as well as a continuous, symmetry-enforced, direct gap at every k-point, one can calculate the $\mathbb{Z}_2$ topological invariant between each pair of bands near the Fermi level by simply analyzing the parity of the wave function at the TRIM (time-reversal invariant momentum) points \cite{fu2007topological}.  This analysis reveals a number of topologically nontrivial crossings between adjacent bands in the region $\pm$ 1\,eV from the Fermi level. For clarity, we will focus on the surface states crossing at the  $\overline{M}$-point here with further analysis presented in the supplemental material \cite{supplemental}. Fig. \ref{fig:topo}(b) presents a close-up of the calculated surface states near the $\overline{M}$ point. The surface states at the $\overline{M}$-point manifest approximately 0.05\,eV above the Fermi energy.  The apparent anisotropy in the calculated surface state dispersions ($\overline{M}$-$\overline{K}$ versus $\overline{M}$-$\overline{\Gamma}$) derives from the direct ``gap" moving up or down in energy depending on direction away from the $\overline{M}$ point. This is not uncommon amongst topological metals \cite{PhysRevB.97.075125,PhysRevLett.124.106402,PhysRevLett.115.036807}.  

Topologically nontrivial surface states close to $E_F$ and the continuous direct gap throughout the Brillouin zone allow the identification of the normal state as a $\mathbb{Z}_2$ topological metal \cite{schoop2015dirac, nayak2017multiple}.  The $T^*$ transition in this compound also suggests that electronic interactions are appreciable in this material.  This transition is accompanied by a subtle change in the derivative of the lattice parameters, cell volume, and associated crystallographic parameters upon crossing $T^*$ \cite{supplemental}. Single crystal x-ray diffraction further shows the formation of a weak superlattice of charge scattering at half-integer reflections (an example shown in the inset of Fig. 2 (a)) \cite{supplemental}.  

The presence of a weak, structural superlattice is suggestive of a secondary structural response to a primary electronic order parameter such as a charge or bond density wave instability. Theoretical studies of partially filled kagome lattices predict a wide array of electronic order parameters \cite{wen2010interaction}.  The metallic nature of CsV$_3$Sb$_5$ and its high degree of covalency makes formal charge assignment imprecise; however, in the ionic limit, the kagome lattice of V-sites would possess one electron per triangle (1/6 filling). Charge density wave (CDW) order with a ($\pi$, $0$) in-plane wave vector consistent with our single crystal x-ray diffraction data is predicted in spinful models and spinless fermion models of interacting electrons in a partially filled kagome lattice.\cite{wen2010interaction}

Nesting across a two-dimensional Fermi surface with an underlying hexagonal motif is also thought to promote the formation of a superconducting state \cite{nandkishore2012chiral}.  Competing density wave instabilities may also arise, and in the present case, scattering along the ($\pi$, $0$) wave vector would connect an enhanced density of states at saddle points near the Fermi energy at the M-points in CsV$_3$Sb$_5$'s band structure.  To our knowledge, this is the first material example hosting the band structure, Fermi energy, and ground state requisite for this theoretical mechanism.   Given the CDW-like instability observed at $T^*$ in this compound, interactions along this wave vector are likely enhanced and may promote a competition between CDW and superconducting instabilities.  Although a structural superlattice exists, ARPES data do not resolve spectral broadening of the Fermi surface across the nested M-points, consistent with the long-range, weak nature of the high temperature density wave order.  Unconventional superconductivity with chiral $d$-wave pairing may emerge in this scenario \cite{PhysRevB.87.115135,nandkishore2012interplay}.

Superconductivity manifest within an electronically two-dimensional kagome lattice is rare unto itself.   While other materials with kagome networks embedded within their lattice structures are known to superconduct (e.g. in certain silicides and borides \cite{rauchschwalbe1984superconductivity, lu2015novel}), all of these examples are inherently three-dimensional both structurally and electronically. CsV$_3$Sb$_5$ seemingly opens a unique opportunity for mapping to models of nesting-driven instabilities emergent within a two-dimensional kagome metal.  Isostructural variants KV$_3$Sb$_5$ and RbV$_3$Sb$_5$ host similar $T^*$ transitions at $80$ K and $104$ K respectively, likely indicative of a similar high-temperature density wave order \cite{ortiz2019new}; however superconductivity has only been observed in CsV$_3$Sb$_5$ to date.  Understanding the interplay between the potentially competing $T^*$ order parameter and the formation of superconductivity across the AV$_3$Sb$_5$ family is an interesting topic for future study. 

The $\mathbb{Z}_2$ topological band structure of CsV$_3$Sb$_5$ may also be of interest for stabilizing the formation of Majorana modes within the vortex cores of a natively proximitized, superconducting surface state. Materials hosting both topologically nontrivial surface states and a native superconducting ground state are uncommon, with relatively few promising candidates identified in FeSe$_{(1-x)}$Te$_x$,\cite{wang2015topological,xu2016topological,wu2016topological,zhang2018observation,machida2019zero} doped-Bi$_2$Se$_3$,\cite{fu2010odd,kriener2011bulk,sasaki2011topological,liu2015superconductivity,du2017superconductivity} and Sn$_{(1-x)}$In$_x$Te\cite{sato2013fermiology,novak2013unusual,polley2016observation}. With relatively light electron-doping (such as Ba substitution), CsV$_3$Sb$_5$ can likely be driven into a regime where such a proximitized topological surface state could be tested.       

In summary, our results demonstrate that kagome metals can serve as a rich arena for exploring the interplay between correlated electron effects and superconductivity within a topologically nontrivial band structure. Our results demonstrate bulk superconductivity with $T_c = 2.5$\,K in single crystals of CsV$_3$Sb$_5$ and classify its normal state as a $\mathbb{Z}_2$ topological metal with multiple topologically nontrivial band crossings in close proximity to the Fermi level.  An anomalous CDW-like transition in the normal state suggests strong correlation effects and an electronic instability that weakly couples to the lattice.  Future studies exploring the relation between this instability and the potential emergence of nesting-driven, unconventional superconductivity on the kagome lattice is motivated by our present results.

\section{\label{sec:ack} Acknowledgments}

S.D.W., R.S., L.B., and B.R.O. acknowledge support from the University of California Santa Barbara Quantum Foundry, funded by the National Science Foundation (NSF DMR-1906325). Research reported here also made use of shared facilities of the UCSB MRSEC (NSF DMR-1720256). B.R.O. and P.M.S. also acknowledge support from the California NanoSystems Institute through the Elings Fellowship program. S.M.L.T has been supported by the National Science Foundation Graduate Research Fellowship Program under Grant No. DGE-1650114. This research used resources of the Advanced Photon Source, a U.S. Department of Energy (DOE) Office of Science User Facility operated for the DOE Office of Science by Argonne National Laboratory under Contract No. DE-AC02-06CH11357. J.-F. H. and Y. H. were supported by the USTC start-up fund. The ARPES measurements were carried out under the user proposal program of SSRL, which is operated by the Office of Basic Energy Sciences, US DOE, under contract No. DE-AC02-76SF00515.  This research used beamline 28-ID-1 of the National Synchrotron Light Source II, a U.S. Department of Energy (DOE) Office of Science User Facility operated for the DOE Office of Science by Brookhaven National Laboratory under Contract No. DE-SC0012704.

\bibliography{CsV3Sb5}

\end{document}